\documentclass[11pt]{article}
\usepackage{amsmath}

\let\ssection=\section
\renewcommand{\section}{\setcounter{equation}{0}\ssection}

\newcommand{\SO}[1]{{\mathop{\rm SO}}({#1})}
\newcommand{\half}{{\scriptstyle{\frac{1}{2}}}}

\newcommand{\IR}{{\bf R}}

\newcommand{\vx}{{\vec x}}

\newcommand{\const}{\mathop{\rm const}\nolimits}
\def\parag{\hfil\break} 
\def\kikezd{\parag\underbar}

\def\vb{{\vec b}}
\def\vv{{\vec v}}
\def\vK{{\vec K}}

\def\vP{{\vec P}}
\def\vQ{{\vec Q}}
\def\vS{\vec{S}}
\def\valpha{\vec{\alpha}}
\def\veta{\vec{\eta}}

\def\vx{{\vec x}}
\def\vX{{\vec X}}

\begin{document}

\setlength{\baselineskip}{16pt}

\title{Mathisson' spinning electron~:
  noncommutative mechanics \& exotic Galilean symmetry,
  66 years ago\\[8pt]
}
\author{
P.~A.~Horv\'athy
\\
Laboratoire de Math\'ematiques et de Physique Th\'eorique\\
Universit\'e de Tours\\
Parc de Grandmont\\
F-37 200 TOURS (France)
}

\date{\today}

\maketitle

\begin{abstract}
     The acceleration-dependent system with noncommuting coordinates,
     proposed by Lukierski, Stichel and Zakrzewski [Ann. Phys. 260,
     224 (1997)]
     is derived as the non-relativistic limit of Mathisson's
     classical electron [Acta Physica Polonica 6, 218 (1937)],
     further discussed by Weyssenhoff and Raabe
     [Acta Physica Polonica 9, 7 (1947)].
     The two-parameter centrally extended Galilean symmetry of
     the model is recovered using elementary methods.
     The relation to Schr\"odinger's Zitternde Elektron
     is indicated.
\end{abstract}

\noindent
Acta Physica Polonica {\bf 34}, 2611 (2003)
[\texttt{hep-th/0303099}].

\section{Introduction}

Non-commutative (quantum) mechanics, where the position
coordinates satisfy
\begin{equation}
	\big\{X_{1}, X_{2}\big\}=\theta,
\label{NCpos}
\end{equation}
has been at the center of recent interest \cite{NCQM}.
In the plane and in the non-relativistic context,
such theories are closely related to the
``exotic'' Galilean symmetry associated with the
two-fold central extension of the planar Galilei
group \cite{exotic}.
A model which provides a physical realization
of this symmetry has been presented by Lukierski,
Stichel and Zakrzewski \cite{LSZ} who considered
the acceleration-dependent Lagrangian
\begin{equation}
     L=
     \frac{m\dot{\vx}^2}{2}
     +
     \frac{\kappa}{2}\,\dot{\vx}\times\ddot{\vx}.
     \label{LSZlag}
\end{equation}

My aim here is to point out that the
model of Lukierski et al. can actually be derived
from that published by Mathisson in `37
\cite{Mathisson}, and further discussed by Weyssenhoff
and Raabe \cite{WeyRaa}.
Not surprisingly, their theory shows interesting analogies also
with Schr\"odinger's {\it Zitternde Elektron} \cite{Schr}.

This Note is dedicated
to the memory of these outstanding physicists who,
with an extreme courage, tried  to continue
their scientific activity
under those terrible years of World War II.

\section{The Mathisson electron}\label{Mathisson}

Two years before the outbreak of World War II,
  Mathisson \cite{Mathisson} proposed to
describe a classical electron with the relativistic equations
\begin{equation}
     \begin{array}{ll}
     m\dot{u}^\alpha+\displaystyle\frac{1}{c^2}S^{\alpha\sigma}
     \ddot{u}_{\sigma}=f^\alpha
     \\[8pt]
     \dot{S}^{\alpha\beta}-
     \displaystyle\frac{1}{c^2}S^{\alpha\sigma}
     \dot{u}_{\sigma}u^\beta
     +
     \displaystyle\frac{1}{c^2}S^{\beta\sigma}\dot{u}_{\sigma}u^\alpha
     =0
     \end{array}
     \label{RMeq}
\end{equation}
where $m$ is the mass, $u^\alpha$ the four-velocity,
$f^\alpha$ the force; the dot means differentiation w.r.t. proper time.
The antisymmetric tensor $S^{\alpha\beta}$ represents the
spin of the electron and is assumed to
satisfy the orthogonality condition
\begin{equation}
     S^{\alpha\beta}u_{\beta}=0.
     \label{spinconstr}
\end{equation}
In the rest frame, the spatial components of $S^{\alpha\beta}$
form therefore a three-vector $\vS$.

In the non-relativistic limit,
$\vS$ becomes a constant of the motion. In the absence of external force,
the motion is [apart of free motion along the direction of $\vS$],
in the plane perpendicular to $\vS$ and satisfies the third-order equation
\begin{equation}
     m\ddot{x}_{i}=-\kappa\epsilon_{ij}\dddot{x_{j}},
     \label{Mateq}
\end{equation}
where the new constant $\kappa$ has been defined by the
Jackiw-Nair Ansatz \cite{JaNa}
\begin{equation}
     s=\kappa c^2,
     \label{JNAnsatz}
\end{equation}
$s=\vert\vS\vert$ being the length of the spin vector.
Eqn. (\ref{Mateq}) is precisely the equation of motion
put forward by of Lukierski et al. \cite{LSZ}.

 From now on we drop the coordinate parallel to
$\vS$ and focus our attention to motion in the plane.

\section{Conserved Quantities}

The equations of motion (\ref{Mateq}) are associated with
the Lagrangian (\ref{LSZlag}).
Then Lukierski et al. derive the conserved quantities
associated to the space-time symmetries
applying the higher-order version of Noether's theorem.
Let us now reproduce their results using elementary methods.

$\bullet$
An obvious first integral of (\ref{Mateq}) is the momentum,
\begin{equation}
     P_{i}=m\dot{x}_{i}+\kappa\epsilon_{ij}\ddot{x}_{j}.
     \label{momentum}
\end{equation}
Eq. (\ref{Mateq}) is in fact $\dot{P}_{i}=0$.

$\bullet$ Multiplying (\ref{Mateq}) by the velocity,
$\dot{\vx}$, yields a total time derivative, where we recognize the
conserved {\it energy},
\begin{equation}
     H=\frac{m\dot{\vx}\strut^2}{2}+\kappa\,\dot{\vx}\times\ddot{\vx}.
     \label{energy}
\end{equation}

$\bullet$ Similarly, taking the vector product of
(\ref{Mateq}) with $\vx$ yields the conserved {\it angular momentum},
\begin{equation}
     J=m\vx\times\dot{\vx}+\frac{\kappa}{2}\dot{\vx}\strut^2
     -\kappa\vx\cdot\ddot{\vx}.
     \label{angmom}
\end{equation}

$\bullet$ A Galilean boost $\vx\to\vx+\vb t$ shifts the momentum
as $\vP\to\vP+m\vb$. A rest frame where the momentum vanishes can
be found, providing us with the conserved {\it boost} vector
\begin{equation}
     K_{i}=mx_{i}-t\big(m\dot{x}_{i}+\kappa\epsilon_{ij}\ddot{x}_{j}\big)
     +\kappa\epsilon_{ij}\dot{x}_{j}.
     \label{CM}
\end{equation}

Somewhat surprisingly, one more conserved quantity can be found.

$\bullet$  the vector product of
(\ref{Mateq}) with the acceleration, $\ddot{\vx}$,
yields the square of the acceleration,
\begin{equation}
     I=\frac{\kappa^3}{2m^2}\big(\ddot{\vx}\big)^2,
     \label{intangmom}
\end{equation}
where a constant factor
has been included for later convenience.

$\bullet$ curiously, multiplying (\ref{Mateq}) by $\dddot{\vx}$ yields
once again the same  quantity, namely
$(m/2)\big(\ddot{\vx}\big)^2=(m/\kappa)^3I$.

The construction of this new quantity reminds one to that
of  angular momentum and of energy.
Its precise origin will be clarified below.

Let us observe that, owing to the conservation
of $I$, $\ddot{\vx}=0$ can be consistently required.
Then the conserved quantities found above
reduce to those of an ``elementary exotic particle'' studied
in \cite{DH}.

\section{Zitterbewegung and center-of-mass
decomposition}\label{NRmotion}

The equation of motion (\ref{Mateq}) is integrated at once.
Putting indeed
\begin{equation}
     Q_{i}=
     -\big(\frac{\kappa}{m}\big)^2\epsilon_{ij}\ddot{x}_{j},
     \label{intcoord}
\end{equation}
eqn. (\ref{Mateq}) becomes
\begin{equation}
     \dot{Q}_{i}=\frac{m}{\kappa}\epsilon_{ij}Q_{j},
     \label{Qeq}
\end{equation}
showing that  the
acceleration rotates uniformly with angular velocity
$m/\kappa$. Putting $Q=Q_{1}~+~iQ_{2}$, $Q(t)=Q_{0}e^{-i(m/\kappa)t}$.
This is plainly consistent with the conservation
of the magnitude of the acceleration, Eq. (\ref{intangmom}). Then
\begin{equation}
     X_{i}=x_{i}+\epsilon_{ij}Q_{j}
     \label{CMcoord}
\end{equation}
moves freely,
\begin{equation}
     \ddot{X}_{i}=0.
     \label{CMmot}
\end{equation}
In conclusion, the motion has been separated into the
free motion of the center of mass coordinate
$\vX$, combined with the ``Zitterbewegung'' [uniform rotation] of the
internal coordinate $\vQ$.

A key feature of Mathisson's electron is that
  the internal variable
$\vQ$ measures in fact the extent of how much
the momentum, $\vP$, differs from [$m$-times] the velocity, $\dot{\vx}$,
\begin{equation}
     \vQ=\frac{\kappa}{m^2}(m\dot{\vx}-\vP).
     \label{velmom}
\end{equation}

Re-writing the conserved quantities in terms of the new
coordinates confirms the above interpretation. In fact,
\begin{equation}
     \begin{array}{ccc}
	\vP&&=m\dot{\vX}\hfill
	\\[3pt]
	H&=H_{CM}+H_{int}\hfill&=
	\displaystyle\frac{m\dot{\vX}^2}{2}-
	\displaystyle\frac{m^3}{2\kappa^2}\vQ^2\hfill
	\\[6pt]
	J&=J_{CM}+J_{int}\hfill&=
	m\vX\times\dot{\vX}+\displaystyle\frac{\kappa}{2}\dot{\vX}^2+
	\displaystyle\frac{m^2}{2\kappa}\vQ^2\hfill
	\\[8pt]
	K_{i}&&=m(X_{i}-\dot{X}_{i}t)+\kappa\epsilon_{ij}X_{j}\hfill
	\\[6pt]
	I&&=\displaystyle\frac{m^2}{2\kappa}\vQ^2.\hfill
     \end{array}
     \label{decompCC}
\end{equation}

Mathisson's electron  is hence a composite system.
Note that in (\ref{decompCC}) the center of mass
behaves precisely as an
elementary exotic particle \cite{DH};
the internal coordinate only contributes to the energy
and the angular momentum. In fact,
$H_{int}=-\displaystyle\frac{m}{\kappa}I$
and $J_{int}=I$.
The  new conserved quantity found in (\ref{intangmom}) is hence
the internal angular momentum and also the internal energy
[which are linked in a $2$-dimensional phase space].

Let us now observe that the equations of motion (\ref{Qeq}-\ref{CMmot})
are consistent with the Poisson structure associated with the
symplectic form
\begin{equation}
     \Omega=\Omega_{CM}+\Omega_{int}=
     dP_{i}\wedge dX_{i}
     +\frac{\kappa}{2m^2}\epsilon_{ij}dP_{i}\wedge dP_{j}
     +\frac{m^2}{\kappa}\epsilon_{ij}
     dQ_{i}\wedge dQ_{j}.
     \label{sympstr}
\end{equation}
The $6$ dimensional phase space is hence the direct sum
of the four-dimensional ``exotic'' phase space
of the center of mass with coordinates $\vX$ and $\vP$,
with the two-dimensional internal phase space of the
$\vQ$, endowed with a canonical symplectic structure.

The Poisson structure can be used to calculate the
algebraic structure of the symmetries.
Consistently with Lukierski et al. \cite{LSZ},
we find that $\vP, H, J, \vK$, supplemented with the
central charges $m$ and $\kappa$,
realize the ``exotic''
[two-fold centrally extended] planar Galilei group.
The  structure relations of
this latter only differ from those of
the usual Galilei group in that
the Poisson bracket of the boost components yields the
``exotic'' central charge,
\begin{equation}
     \big\{K_{1},K_{2}\big\}=\kappa.
     \label{exoCR}
\end{equation}
Similarly, the center-of-mass coordinates have a nonvanishing
Poisson bracket,
\begin{equation}
     \big\{X_{1},X_{2}\big\}=\frac{\kappa}{m^2},
     \qquad
     \big\{Q_{1},Q_{2}\big\}=-\frac{\kappa}{m^2}.
     \label{NCCR}
\end{equation}
Both the center-of-mass and the internal coordinates are hence
noncommuting, cf. (\ref{NCpos}) with $\theta=(\kappa/m^2)$
[while the original coordinates $x_{i}$ commute].
This is similar to what happens
in the Landau problem where the guiding center coordinates are
noncommuting, with $\theta=1/eB$.

The additional conserved quantity $I$ in
(\ref{intangmom}) is actually associated with
the {\it internal symmetries}  of the system.
The translations and boosts form indeed an invariant
subgroup $K$ of the Galilei group.
The quotient $G/K$, which consists of
rotations and time translations, is hence a group that can be made to
act separately on the center-of-mass and the internal space.
We can, e. g., rotate
the internal coordinate $\vQ$ alone and leave the center-of-mass
coordinate $\vX$ fixed. This is plainly a symmetry,
and the associated conserved quantity is the
internal angular momentum $J_{int}=I$. (A physical rotation
moves both the external and internal coordinates, yielding
the total angular momentum in (\ref{decompCC})).
The internal energy arises in a similar way.
In conclusion, the non-relativistic limit of the
Mathisson electron admits the direct product
of the ``exotic'' Galilei group with the internal
rotations and time translations, $\SO2\times \IR$, as symmetry.
Here the action of the Galilei group is transitive on the
submanifolds $I=\const$ i. e., $\vQ^2=\const$.

The same statement is valid for any composite
nonrelativistic system, i. e. one
upon which the Galilei group acts by symmetries but not
transitively \cite{SSD}.

\section{Relation to Schr\"odinger's Zitternde Elektron}\label{Zitter}

The results of Section \ref{NRmotion} remind
those  Schr\"odinger derived in his original paper on
Zitterbewegung \cite{Schr}.
Schr\"odinger starts in fact with the Dirac Hamiltonian
\begin{equation}
     H=c\valpha\cdot\vP+m^2c^2\beta
     \label{Dham}
\end{equation}
where $\valpha$ and $\beta$ denote the usual Dirac matrices.
In the Heisenberg picture, the operators satisfy
\begin{equation}
\frac{d\vP}{dt}=0,
\qquad
\frac{dH}{dt}=0
\qquad
\frac{d\vx}{dt}=c\valpha.
\label{Heq}
\end{equation}
The last equation can be rewritten as
$
-i\frac{d\veta}{dt}=2H\veta,
$
($\hbar=1$), where
$
\veta=\valpha-cH^{-1}\vP.
$
This can be integrated as
$\veta(t)=e^{2iHt}\veta_{0}=\veta_{0}e^{-2iHt}$,
where $\veta_{0}$ is a constant operator. Hence
\begin{eqnarray*}
\frac{d\vx}{dt}=c^2H^{-1}\vP+c\veta_{0}e^{-2iHt},
\end{eqnarray*}
which can again be integrated to yield
\begin{equation}
     \vx(t)=
     \big\{\vX_{0}+c^2H^{-1}\vP t\big\}
     +\half ic\vec{\eta}_{0}H^{-1}e^{-2iHt}
     \label{zitter}
\end{equation}
where $\vX_{0}$ is a constant operator.
The structure is clearly the same as in
(\ref{CMcoord}), with the operator
\begin{equation}
     \vX(t)=\vX_{0}+c^2H^{-1}\vP t
\end{equation}
representing the
freely moving center-of-mass, and the second term describing
the internal Zitterbewegung. The precise relation is more subtle,
though.
Intuitively, dropping the third component and working in the plane,
putting $s=1/2$ and $s/c^2\simeq \kappa$ [which would require the
spin to diverge as $c\to\infty$ rather then remain a constant], setting
$c\valpha\simeq \dot{\vx}$ and replacing $H\simeq mc^2$,
would transform (\ref{zitter}) formally into (\ref{CMcoord}). In fact,
\begin{eqnarray}
     \vX(t)\simeq \vX_{0}+\frac{\vP}{m}t,
     \qquad
     mc\veta\simeq\frac{m^2}{\kappa}\vQ,
     \qquad
     e^{-i2Ht}=e^{-i(H/s)t}\simeq e^{-i(m/\kappa)t}.
\end{eqnarray}
Note that
\begin{equation}
     mc\veta=mc\valpha-mc^2H^{-1}\vP\simeq m\dot{\vx}-\vP.
\end{equation}
consistently with (\ref{velmom}).

A distinctive feature of Schr\"odinger's Zitternde Elektron
is that the center-of-mass
coordinates satisfy the nontrivial commutation relation
\begin{equation}
     \big[X_{i},X_{j}]=-ic^2E^{-2}\epsilon_{ijk}S_{k},
\end{equation}
where $E=c\sqrt{\vP^2+m^2c^2}$ and $\vS=-(i/4)\valpha\times\valpha$
is the spin operator. If we assume that
the spin is polarized in the
third direction, $S_{3}=-s$, and we consider the non-relativistic
limit $E\simeq mc^2+\vP^2/2m$ together with the Ansatz
(\ref{JNAnsatz}), we find for the planar components
\begin{equation}
     \big[X_{1},X_{2}]\simeq i\frac{s}{c^2m^2}=i\frac{\kappa}{m^2}
\end{equation}
cf. (\ref{NCpos}) with $\theta=\kappa/m^2$.
Let us remark that our procedure here is in fact the quantum version of
the subtle non-relativistic limit proposed by Jackiw and Nair
\cite{JaNa}.

\section{The relativistic description of Weyssenhoff and Raabe}

Mathisson's classical electron was further elaborated
by Weyssenhoff and Raabe in a paper
published after the War \cite{WeyRaa}. They posit the equations
\begin{equation}
     \begin{array}{ll}
     \dot{p}^\alpha=0,\qquad
     p^\alpha=mu^\alpha+
     \displaystyle\frac{1}{c^2}S^{\alpha\beta}\dot{u}_{\beta},
     \\[8pt]
     \dot{S}^{\alpha\beta}=p^\alpha u^{\beta}-p^\beta u^{\alpha},
     \\[6pt]
     S^{\alpha\sigma}u_{\sigma}=0,
     \end{array}
     \label{WeRaeq}
\end{equation}
where
$
m=-\frac{1}{c^2}u_{\beta}p^\beta.
$
Eliminating $p^\alpha$ yields the relativistic
Mathisson equations (\ref{RMeq})
once again. Eqns. (\ref{WeRaeq}) imply that $m$ is
constant of the motion, $\dot{m}=0$, identified as the rest-mass
of the particle. $S_{\alpha\beta}S^{\alpha\beta}=s^2$
is also a constant of the motion. They also
observe that, owing to $\dot{p}^\alpha=0$, the quantity
$M$ defined by $p_{\alpha}p^{\alpha}=M^2c^2$ is another constant of
the motion.
It is worth noting that the position satisfies again
a third-order equation analogous to (\ref{Mateq}), namely
\begin{equation}
     m\ddot{x}_{\alpha}=-\frac{1}{c^2}S_{\alpha\sigma}\dddot{x}_{\sigma}.
     \label{WR3}
\end{equation}

Then Weyssenhoff and Raabe proceed to integrate the free relativistic
equations of motion.
In a suitable inertial frame (called the proper system)
the spatial components, $P_{i}$, of the
vector $p^\alpha$ can be made to vanish,
so that its time component is $Mc$.
In this frame $\vS$ is constant. The mass is
$m=M/\sqrt{1-(\vv/c)^2}$ where $v_{i}=u_{i}\sqrt{1-(\vv/c)^2}$
denotes the three-velocity. Hence the time component of the four-velocity
is also constant so that the four-acceleration is proportional
to the three-acceleration, ${\vec a}=d^2\vx/dt^2$.
Transforming from proper time to $t$, $\vP=0$ reduces finally to
\begin{equation}
     M\vv+\frac{1}{c^2}\vS\times{\vec a}=0.
\end{equation}
The particle moves hence along a circle
in the plane perpendicular to $\vS$,
with uniform angular velocity
\begin{equation}
     \frac{mc^2}{s}\Big(1-\frac{\vv^2}{c^2}\Big).
\end{equation}
In a general Lorentz frame, the motion is a superposition of
such a motion with a uniform translation.

Our clue is to observe that in the
non-relativistic limit these formul{\ae} reduce,
with the Jackiw-Nair Ansatz $s=\kappa c^2$
cf. (\ref{JNAnsatz}),
to those we derived in Section \ref{NRmotion}.

It is worth mentionning that the equations of
Weyssenhoff and Raabe have again and again re-emerged
in the course of the years. Consider, for example,
(\ref{WeRaeq}) in{\it five} dimensions and for
$s=\half$. Multiplication of $p^\beta$ with $S_{\alpha\beta}$
allows us to express the five-vector $\dot{u}_{\alpha}$
as
\begin{equation}
     \dot{u}_{\alpha}=\frac{4}{c^2}S_{\alpha\sigma}p^\sigma
\end{equation}
which, together with the remaining relations in (\ref{WeRaeq})
and the constraint $u_{\alpha}u^{\alpha}=1$, are
precisely the equations proposed by Barut and Zanghi
\cite{Barut} as a ``Kaluza-Klein'' description of
a classical Dirac electron.

\section{Conclusion}

In this Note we have shown that the non-relativistic limit of
Mathisson's classical spinning electron yields
the acceleration-dependent model of Lukierski et al. \cite{LSZ}.
This latter has non-commuting coordinates and
  realizes the ``exotic'' Galilean symmetry.

Our results confirm once again the relation between the relativistic
spin and the non-relativistic ``exotic'' structure, advocated
  by Jackiw and Nair \cite{JaNa}.
Their rule (\ref{JNAnsatz}) is,
however, a rather strange one, since it requires
the spin to diverge as $c\to\infty$
so that $s/c^2$ remains finite.
For this reason, the use of a Dirac equation valid for the
fixed velue $s=\half$
[as in Section \ref{Zitter} above] is clearly illegitimate, and should
be replaced by some anyon equation,
  valid for any real spin $s$ \cite{aneq}.

Another intriguing feature of this procedure is the following.
While the relativistic model is  associated with an
irreducible representation of the Poincar\'e group,
its dequantized \& non-relativistic limit, namely
the model of Lukierski et al., only carries a
reducible representation of the Galilei group~:
irreducibility is lost in the procedure.

A final remark concerns the spin constraint
(\ref{spinconstr}) which appears to lie at
the very root of the Zitterbewegung. Trading it
for
\begin{equation}
     S^{\alpha\beta}p_{\beta}=0
     \label{Pspinconstr}
\end{equation}
would in fact eliminate the Zitterbewegung altogether and
lead to models of the type discussed in \cite{Dixon}.

\kikezd{\bf Acknowledgement.}
I am indebted to Professor
J. Lukierski for sending me
copies of those old Acta Physica Polonica  papers, and
also to Professor A. Staruszkiewicz
who provided me with some biographical data.



\begin{thebibliography}{99}
\bibitem{NCQM}
It is impossible to provide a complete list of references.
See, e. g.,
V.~P.~Nair and A.~P.~Polychronakos,
{\it Quantum mechanics on the noncommutative plane and sphere}.
{\sl  Phys. Lett}. {\bf B 505}, 267 (2001);
J.~Gamboa, M.~Loewe, F.~M\'endez, and J.~C.~Rojas,
{\it The Landau problem in noncommutative Quantum Mechanics};
  S.~Bellucci, A. Nersessian, and C. Sochichiu,
{\it Two phases of the noncommutative quantum mechanics}.
{\sl Phys. Lett.} {\bf B522}, 345 (2001), etc.

\bibitem{exotic}
J.-M.~L\'evy-Leblond,
in {\it Group Theory and Applications} (Loebl Ed.),
{\bf II}, Acad. Press, New York, p. 222 (1972);
A.~Ballesteros, N. Gadella, and M.~del Olmo,
{\it Moyal quantization of\ $2+1$ dimensional
   Galilean systems}.
  {\sl Journ. Math. Phys.} {\bf 33}, 3379 (1992);
Y.~Brihaye, C.~Gonera, S.~Giller and P.~Kosi\'nski,
  {\it Galilean invariance in $2+1$ dimensions.}
\texttt {hep-th/9503046} (unpublished);
D.~R.~Grigore,
  {\it Transitive symplectic manifolds in $1+2$ dimensions}.
  {\sl Journ. Math. Phys.} {\bf 37}, 240 (1996);
   {\it The projective unitary irreducible representations of the
  Galilei group in $1+2$ dimensions}. {\sl ibid}. {\bf 37}, 460 (1996).

\bibitem{LSZ}
J.~Lukierski, P.~C.~Stichel, W.~J.~Zakrzewski,
   {\it Galilean-invariant $(2+1)$-dimensional models with a
     Chern-Simons-like term and $d=2$ noncommutative geometry}.
  {\sl Annals of Physics}  (N. Y.) {\bf 260}, 224 (1997).
The model is further discussed in
P.~A. Horv\'athy and M.~S.~Plyushchay,
{\it Non-relativistic anyons, exotic Galilean symmetry and the
non-commutative plane}.
{\sl JHEP} {\bf 06} (2002) 033; 
  J.~Lukierski, P.~C.~Stichel, W.~J.~Zakrzewski,
   {\it Noncommutative planar particle dynamics
   with gauge interaction}.
  {\sl Annals of Physics}  (N. Y.) (in press)
[\texttt{hep-th/0207149}].

\bibitem{Mathisson}
M.~Mathisson,
{\it Das Zitternde Elektron und seine Dynamik}.
{\sl Acta Physica Polonica} {\bf 6}, 218-227 (1937).
Myron Mathisson (1897-1940), of Jewish origin,
taught mathematical physics at Warsaw University
as a {\it Privatdozent}. He also worked in Krak\'ow,
benefitting of a kind of ``private scholarship''
created for him by Weyssenhoff.
Then he spent one year in Kazan, in the Soviet Union.
In 1939 he escaped to Britain, where he died.
He was remembered by
Dirac in the Obituary reproduced below, published
in  {\sl Nature} {\bf 146}, 613 (1940)~:
``The death of Dr. Myron Mathisson on September 13 at the early
age of fourty-three has cut short an interesting line of research.
Mathisson had been engaged for many years in studying the general
dynamical laws governing the motion of a particle,
with possibly a spin or a moment, in a gravitational or
electromagnetic field, and had developed a powerful method of his own
for passing from field equations to particle equations. The subject
is of particular interest at the present time, as it has now become
clear that quantum mechanics cannot solve the difficulties that arise
in connexion with the interaction of point particles with fields,
and a deeper classical analysis of the problem is needed.
It is much to be regretted that Mathisson's death has occured
before the relations between his method and those of other workers
on the subject have been completely elucidated.

Mathisson carried out his work at the Universities of Warsaw and Kazan
and at an institute which he started in Cracow, and, since the spring
of 1939, at Cambridge.''

[Sources: A short history of Theoretical Physics at Hoza 69 \dots and 
personal communication of
Prof. A. Staruszkiewicz.]

\bibitem{WeyRaa}
J.~Weyssenhoff and A. Raabe,
{\it Relativistic dynamics of spin fluids and spin particles}.
{\sl Acta Physica Polonica} {\bf 9}, 7-18
(1947). Let us also record the footnote written
by Weyssenhoff. ``Presented at a meeting of the
Cracow Section of the Polish Physical Society on February 28, 1945.
[\dots] Most of the results were subject of a lecture at a
secret meeting of physicists at Prof. Pie\'nkowski's home
in Warsaw, October 1942.

Mr. Raabe was a highly gifted young physicist with whom I outlined in
all its main features the contents of this paper in 1940/41 in Lw\'ow.
We tried to pursue our work in 1942 in Cracow, but unfortunately
in June 1942 Mr. Raabe fell victim of a man-hunt in the streets of
Cracow; he died four months later in the German concentration camp
O\'swi\c ecim [Auschwitz].''

Jan Weyssenhoff (1889-1972) came from a prominent
Baltic-German aristocratic family, which remained
Catholic and become Polish in the XVIIth century.
He was a gentleman in the old sense of the word, who
used his personal fortune and his wealthy
friends to help other colleagues.
His father was a succesful writer. His mother came from a
very wealthy Jewish banking family which owned, among other things,
the Warsaw-Vienna railway.

Weyssenhoff studied in Krak\'ow and in Z\"urich,
where he also met Einstein,
who refers to him in his work on Brownian Motion
[available in Dover Publications].
He was also interested in the Hall effect and wrote his Ph. D.
on the theory of paramagnetism. He returned to his country in 1919.
   He got involved in the study of relativistic spinning particles
   and fluids in 1937.
  Between 1939 and 1941 he worked at the
Polytechnical University in  Lw\'ow, occupied by the Soviet
army and attached to Ukrain.
In 1942 he returned to Krak\'ow, and was
followed by Raabe, who lived in his flat and
whom he helped also to get documents, e. g., a ``Kennkarte''.

Professor Weyssenhoff also organized secret seminars on physics in his home.
Unlike his young collaborator, he survived to the war
and continued his scientific work until
his death in Krak\'ow,  in 1972.

\bibitem{Schr}
E.~Schr\"odinger,
{\sl Sitzungsber. Preuss. Akad. Wiss.
Phys. Math. Kl}. {\bf 24}, 418 (1930). A summary of
Schr\"odinger's original method is given by
A.~O.~Barut and A.~J.~Bracken,
{\it Zitterbewegung and the internal geometry of the
electron}.
{\sl Phys. Rev.} {\bf D23}, 2454 (1981).

\bibitem{JaNa}
R.~Jackiw and V.~P.~Nair,
{\it Anyon spin and the exotic central extension of the
   planar Galilei group}.
{\sl Phys. Lett.} {\bf B 480}, 237 (2000).

\bibitem{DH}
C.~Duval and P.~A.~Horv\'athy,
  {\it The exotic Galilei group and the ``Peierls substitution''}.
{\sl Phys. Lett.} {\bf B 479}, 284 (2000);
  {\it Exotic galilean symmetry in the non-commutative plane, and the
Hall effect}.
{\sl Journ. Phys}. {\bf A 34}, 10097 (2001).

\bibitem{SSD}
J.-M.~Souriau,
{\it Structure des syst\`emes dynamiques}.
Dunod: Paris (1970).

\bibitem{Barut}
A.~O.~Barut,
{\it Classical model of the Dirac electron}.
{\sl Phys. Rev. Lett}. {\bf 52}, 2009 (1984).

\bibitem{aneq}
An incomplete list of references includes
M.~S.~Plyushchay,
{\it Relativistic model for anyon}.
{\sl Phys. Lett}. {\bf B248}, 107 (1990);
R.~Jackiw and V.~P.~Nair,
{\it Relativistic wave equation for anyons}.
{\sl Phys. Rev.} {\bf D43}, 1933 (1990);
S.~Ghosh,
{\it Spinning particles in $2+1$ dimensions}.
{\sl Phys. Lett}. {\bf B338}, 235 (1994).


\bibitem{Dixon}
W.~G.~Dixon,
  {\it On a classical theory of charged particles with spin and the
  classical limit of the Dirac equation}.
  {\sl Il Nuovo Cimento} {\bf 38}, 1616 (1965);
J.-M.~Souriau,
{\it Mod\`ele de particule \`a spin dans le champ
\'electromagn\'etique et gravitationnel}.
  {\sl Ann. Inst. Henri Poincar\'e}, {\bf 20 A}, 315 (1974);
Ch.~Duval,
{\it The general relativistic Dirac-Pauli particle:
an underlying classical model}.
{\sl Ann. Inst. Henri Poincar\'e}, {\bf 25 A}, 345 (1976).


\end{thebibliography}
\end{document}